\documentclass[twocolumn,showpacs,prl]{revtex4}
\usepackage{graphicx}
\usepackage{dcolumn}
\usepackage{bm}

\begin{document}

\title{Superfluid shells for trapped fermions with mass and population imbalance}
\author{G.-D. Lin, W. Yi and L.-M. Duan}
\address{FOCUS center and MCTP, Department of Physics, University of Michigan, Ann
Arbor, MI 48109}

\begin{abstract}
We map out the phase diagram of strongly interacting fermions in a
potential trap with mass and population imbalance between the two
spin components. As a unique feature distinctively different from
the equal-mass case, we show that the superfluid here forms a
shell structure which is not simply connected in space. Different
types of normal states occupy the trap regions inside and outside
this superfluid shell. We calculate the atomic density profiles,
which provide an experimental signature for the superfluid shell
structure.
\end{abstract}

\maketitle

The recent experimental advance on superfluidity in polarized
ultracold Fermi gas has raised strong interest in studying the
phase configuration of this system
\cite{1,2,3,4,5,6,7,wei1,8,9,10}. The experiments have suggested a
phase separation picture with a superfluid core surrounded by a
shell of normal gas \cite{1,2,3,4}. This picture has been
confirmed also by a number of theoretical calculations of the
atomic density profiles in the trap \cite {wei1,8,9,10}. As
Feshbach resonances between different atomic species with unequal
mass have been reported \cite{FR}, the next step is to consider
the properties of strongly interacting Fermi gas with both mass
and population imbalance between the two spin components. There
have been a few recent theoretical works in this direction
\cite{11,12}, with focus on the properties of a homogeneous gas.

In this work, we consider the properties of a \textit{trapped}
strongly interacting Fermi gas with mass and population imbalance.
We map out its zero temperature phase diagram in a harmonic trap
(generally anisotropic) as a function of few universal parameters.
Compared with the equal-mass case, the two-specie Fermi mixture
($^{6}$Li-$^{40}$K mixture for instance) shows a very different
picture of phase separation: it supports a superfluid shell state
in the intermediate trap region, with normal gases of different
characters filling the center and the edge of the trap. This
unusual phase separation picture with non-monotonic superfluid
order parameter in space only occurs for trapped fermions with
unequal mass. We provide an intuitive explanation for the
phenomenon, and show how to detect it by measuring the atomic
density profiles. This superfluid shell state is not simply
connected in space, so it may support interesting vortex structure
under rotation of the trap.

Strongly interacting Fermi gas near a wide Feshbach resonance can
be well described by the following single-channel Hamiltonian:
\begin{eqnarray}
H &=&\sum_{\mathbf{k},\sigma }\left( \epsilon _{\mathbf{k}\sigma }-\mu
_{\sigma }\right) a_{\mathbf{k,}\sigma }^{\dag }a_{\mathbf{k,}\sigma } \\
&+&\left( U/\mathcal{V}\right) \sum_{\mathbf{q},\mathbf{k},\mathbf{k^{\prime
}}}a_{\mathbf{q}/2+\mathbf{k},\uparrow }^{\dag }a_{\mathbf{q}/2-\mathbf{k}%
,\downarrow }^{\dag }a_{\mathbf{q}/2-\mathbf{k^{\prime }},\downarrow }a_{%
\mathbf{q}/2+\mathbf{k^{\prime }},\uparrow }  \nonumber
\end{eqnarray}
where $\epsilon _{\mathbf{k}\sigma }=\mathbf{k}^{2}/(2m_{\sigma })$ with $%
m_{\sigma }$ denoting the mass of species-$\sigma $ (or called spin $\sigma $%
, with $\sigma $ $=\uparrow ,\downarrow $ and $\hbar=1$),
$\mathcal{V}$ is the quantization volume, and
$a_{\mathbf{k},\sigma }^{\dag }$ is the fermionic creation
operator for the $\mathbf{k}\sigma $ mode. The bare atom-atom
interaction rate $U$ is connected with the physical scattering
length $a_{s}$
through the renormalization relation $1/U=1/U_{p}-(1/\mathcal{V})\sum_{%
\mathbf{k}}{1/(2\epsilon _{\mathbf{k}})}$ with $U_{p}=4\pi a_{s}/(2m_{r})$ (
$m_{r}=m_{\uparrow }m_{\downarrow }/(m_{\uparrow }+m_{\downarrow })$ is the
reduced mass). We take the local density approximation so that $\mu
_{\uparrow }=\mu _{\mathbf{r}}+h,$ $\mu _{\downarrow }=\mu _{\mathbf{r}}-h$,
$\mu _{\mathbf{r}}=\mu -V\left( \mathbf{r}\right) $, where $V\left( \mathbf{r%
}\right) =\sum_{i=x,y,z}\beta _{i}\mathbf{r}_{i}^{2}/2$ is the harmonic
trapping potential, which, without loss of generality, has been assumed to
be the same for the two components. The chemical potential $\mu $ at the
trap center and the chemical potential imbalance $h$ are determined from the
total atom number $N=N_{\uparrow }+N_{\downarrow }$ and the population
imbalance $p=\left( N_{\uparrow }-N_{\downarrow }\right) /N$ through the
number equations below.

Our calculation method is based on a simple extension of the formalism
described in Ref. \cite{wei1} to the unequal mass case. Under the mean-field
approximation, the thermodynamical potential $\Omega =-T\ln [$tr$(e^{-H/T})]$
($T$ is temperature) can be written as \cite{note1}
\begin{eqnarray}
\Omega  &=&-\mathcal{V}\left| \Delta \right| ^{2}/U+\sum_{\mathbf{k}}\left[
\epsilon _{\mathbf{k}\downarrow }-\mu _{\downarrow }-E_{\mathbf{k}\downarrow
}\right]  \\
&&-T\sum_{\mathbf{k}}\ln \left[ \left( 1+e^{-E_{\mathbf{k}\uparrow
}/T}\right) \left( 1+e^{-E_{\mathbf{k}\downarrow }/T}\right) \right]
\nonumber
\end{eqnarray}
where $E_{\mathbf{k}\uparrow ,\downarrow }=E_{\mathbf{k}}\mp (h+\alpha
\epsilon _{\mathbf{k}r})$ with $E_{\mathbf{k}}\equiv \sqrt{(\epsilon _{%
\mathbf{k}r}-\mu _{\mathbf{r}})^{2}+\Delta ^{2}},$ $\epsilon _{\mathbf{k}%
r}\equiv \mathbf{k}^{2}/(4m_{r})$, and $\alpha \equiv (m_{\uparrow
}-m_{\downarrow })/(m_{\uparrow }+m_{\downarrow })$. If one of the
energies $E_{\pm ,\mathbf{k}}$ has one or two zero(s) in
$\mathbf{k}$-space, it signals the presence of the type-I or
type-II breached-pair (Sarma) states \cite{5,7,wei1,wei2} (called
the BP1 or BP2 states, respectively). The BP states represent a
spatially homogeneous superfluid, but they differ from the
conventional BCS\ states by a phase separation in the momentum
space and by a topological change of the Fermi surface for the
excess fermions. To determine the phase at different
trap regions, we search for the global minimum of the thermo-potential $%
\Omega $ with respect to the variational parameter $\Delta $ instead of
using the gap equation $\partial \Omega /\partial \Delta =0$, as the latter
may give unstable or metastable phases as remarked in Ref. \cite{wei1}. The $%
\mu $ and $h$ are determined from the two number equations $N_{\sigma }=\int
d^{3}\mathbf{r}n_{\mathbf{r}\sigma }$ integrated over the trap. The local
atomic density $n_{\mathbf{r}\sigma }$, derived from the thermodynamical
potential $\Omega $ as $\partial \Omega /\partial \mu _{\sigma }=-n_{\mathbf{%
r}\sigma }\mathcal{V}$, has the expression

\begin{equation}
n_{\mathbf{r}\sigma }=\frac{1}{\mathcal{V}}\sum_{\mathbf{k}}[u_{\mathbf{k}%
}^{2}f(E_{\mathbf{k},\sigma })+v_{\mathbf{k}}^{2}f(-E_{\mathbf{k},-\sigma
})],
\end{equation}
where the parameters $u_{\mathbf{k}}^{2}=(E_{\mathbf{k}}+(\epsilon _{\mathbf{%
k}r}-\mu _{\mathbf{r}}))/\left( 2E_{\mathbf{k}}\right) $, $v_{\mathbf{k}%
}^{2}=(E_{\mathbf{k}}-(\epsilon _{\mathbf{k}r}-\mu
_{\mathbf{r}}))/\left( 2E_{\mathbf{k}}\right) $, and the Fermi
distribution $f(E)\equiv 1/\left( 1+e^{E/T}\right) $. For
convenience, we take $-\uparrow =\downarrow $ and vice versa. The
above mean-field formalism is also identical to the $G_{0}G$
diagram scheme if we interpret $\Delta $ at finite temperature as
the total gap which includes contributions from both the order
parameter and the pseudogap associated with the pair fluctuations
\cite{wei1,10}.

We calculate the phase diagram of the trapped fermions in terms of
several dimensionless universal parameters. For that purpose, the
unit of energy is chosen to be the Fermi energy $E_{F}$ at the
center of the trap for $N$ non-interacting fermions with an
effective mass of $2m_{r}$ and with equal population for the two
components. Under the local density approximation,
one finds $E_{F}=(3N\sqrt{\beta _{x}\beta _{y}\beta _{z}})^{1/3}/\sqrt{2m_{r}%
}$ from this definition. The trapping potential $V(\mathbf{r)}$ has the
dimensionless form $V(\mathbf{r)}/E_{F}=\sum_{i}\widetilde{r}_{i}^{2}$,
where the normalized coordinates $\widetilde{r}_{i}\equiv r_{i}/R_{i}$ and $%
R_{i}\equiv \sqrt{2E_{F}/\beta _{i}}$ is the Thomas-Fermi radius along the $i
$th direction. The momentum $\mathbf{k}$ and the temperature $T$ are
measured in the units of $k_{F}\equiv \sqrt{2(2m_{r})E_{F}}$ and $%
T_{F}\equiv E_{F}/k_{B}$, respectively. The system properties then
only depend on four dimensionless parameters $k_{F}a_{s}$,
$T/T_{F},$ the population imbalance $p$, and the mass mismatch
$\alpha $. In the following calculation, we take $\alpha =0.74$
corresponding to the $^{40}$K-$^{6}$Li mixture, as the latter is
the most likely experimental system for the two-specie Fermi gas.

\begin{figure}[tbp]
\includegraphics[width=8cm,height=9cm]{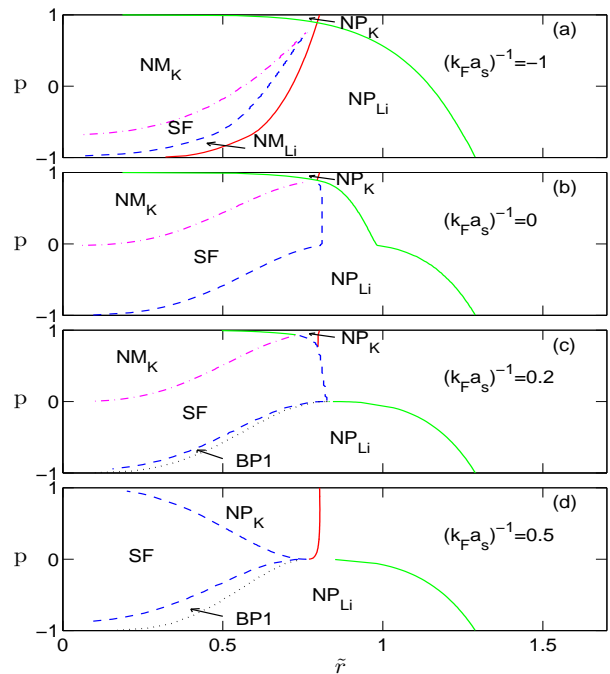}
\caption[Fig.1]{(color online) The zero temperature phase diagrams
for trapped $^{6}$Li-$^{40}$K mixture near a Feshbach resonance. The
phases include the BCS superfluid state (SF), the breached pair
phase of type 1 (BP1), the normal mixture (NM$_{\text{K}}$ or
NM$_{\text{Li}}$, with $^{40}$K or $^{6}$Li in excess,
respectively), and the normal polarized (single-component) states
(NP$_{\text{K}}$ or NP$_{\text{Li}}$). The phases are shown versus
the population imbalance $p$ and the normalized trap radius
$\widetilde{r}$. }
\end{figure}

In Fig. 1, we map out the zero-temperature phase diagrams for
trapped fermions as a function of the population imbalance $p$ at
several characteristic interaction strengths $k_{F}a_{s}$. The
system shows a rich picture of phase separation in the trap.
First, on the BCS\ side of resonance with $\left(
k_{F}a_{s}\right) ^{-1}=-1$ (Fig. 1(a)), even in the equal
population case ($p=0$), we cross four different phases from the
trap center to the edge. The trap center is occupied by a normal
mixture state with the heavy fermions ($^{40}$K) in excess
(denoted as NM$_{\text{K}}$), which is surrounded by a shell of
BCS\ superfluid phase (denoted as SF). Further out, there is a
shell of a normal mixture but now with the light fermions
($^{6}$Li) in excess (denoted as NM$_{\text{Li}}$). The trap edge
is occupied by the single component normal gas of light fermions
(denote as NP$_{\text{Li}}$). The system behavior becomes
significantly more complicated compared with the equal mass case
\cite{wei1,8,9,10}, where instead of several phases, there is only
one superfluid phase over the whole trap in the corresponding
configuration. Note that under the local density approximation, as one
moves out from the trap center, the chemical potential
monotonically decreases, while the superfluid order parameter is
apparently not a monotonic function in this case. The superfluid
only occurs in an intermediate shell. We will give some
explanations for this unusual phenomenon later. Continuing with
the phase diagram, if we increase the population of the heavy
fermions, the central NM$_{\text{K}}$ region grows, while the SF,
NM$_{\text{Li}}$, and the NP$_{\text{Li}}$ phase regions shrink
and finally all disappear at a critical population imbalance.
After that, the trap edge is occupied by a single component normal
gas of heavy fermions (denoted as NP$_{\text{k}}$). If we increase
the population of light fermions, the reverse happens. The central
NM$_{\text{K}}$ region shrinks and finally disappear at a critical
population imbalance, and the superfluid shell evolves into a
superfluid core.

On resonance (Fig. 1(b) with $\left( k_{F}a_{s}\right) ^{-1}=0$),
the superfluid phase region gets significantly larger. The normal
mixture region NM$_{\text{Li}}$ at the intermediate shell
completely disappears, which is quite different from the equal
mass case where there is always such a mixed shell \cite{wei1}.
Moving on to the BEC side of resonance with $\left(
k_{F}a_{s}\right) ^{-1}=0.2$ (Fig. 1(c)), the\ type-I breached
pair phase (BP1) appears at an intermediate shell between the SF\
and NP$_{\text{Li}}$ phases, but only when the light fermions are
in excess. Compared with the equal mass case \cite{wei1}, the
critical $k_{F}a_{s}$ for the  appearance of the BP1 state is
significantly shifted towards the resonance point. Another notable
feature at this $\left( k_{F}a_{s}\right) ^{-1}$ is that when the
heavy fermions are in excess, the superfluid shell is not
surrounded a normal gas anymore. All the normal components are
pushed to the central core. Further on to the BEC side with
$\left( k_{F}a_{s}\right) ^{-1}=0.5$ (Fig. 1(d)), the normal
mixture at the trap center finally disappears at all population
imbalance, and we resume the picture of a superfluid core
surrounded by a shell of a normal gas. The BP1 phase region at the
intermediate shell grows as one expects, but again it only shows
up when the majority is in the light fermions.

A remarkable feature from the above phase diagrams is that the
superfluid forms a shell structure in space, which separates
different types of normal states at the trap center and the edge.
This feature is qualitatively different from the equal mass case.
Now we would like to understand in more detail how this feature
shows up. We know that the phase is determined by the
global minimum of the thermo-potential $\Omega $ as a function of the gap $%
\Delta $, under certain values of the chemical potentials $\mu _{\mathbf{r}}$
and $h$ at the trap position $\mathbf{r}$. As one moves out from the trap
center, $\mu _{\mathbf{r}}$ monotonically decreases as $\mu _{\mathbf{r}%
}=\mu -\mathbf{\widetilde{r}}^{2}$ (in the unit of $E_{F}$) while $h$
remains the same. In Fig. 2(a), we show $\Omega $ as a function of
$\Delta $ at several different values of the normalized radius
$\widetilde{r}$ (thus with different $\mu _{\mathbf{r}}$). The values
of $h$ and the central $\mu $ are taken to be the typical ones for
which there is a superfluid shell structure in the phase diagram.
The potential $\Omega $ typically has a double-well structure. At
the trap center (the lowest curve with $\widetilde{r}=0$), the trivial
well with $\Delta =0$ is deeper which corresponds to a normal
state. As one moves out, both wells are lifted, but with different
speeds. At a lower critical value of $\widetilde{r}$ (which is $0.38$
for the configuration in Fig. 2(a)), the two wells become equally
deep. Above this value, the global minimum jumps to the nontrivial
well with $\Delta \neq 0$, which signals a first-order phase
transition to the superfluid state. As one moves further out, the
nontrivial well approaches the trivial well; and at an upper
critical value of $\widetilde{r}$ (which is $0.81$ in Fig. 2(a)), the
two wells merge, which signals a second-order phase transition
from the superfluid to the normal state. Hence the potential
$\Omega $ varies non-monotonically with $\mu _{\mathbf{r}}$, which
leads to the superfluid shell state only at the intermediate
region.

The above picture is established from the calculation of $\Omega
$. We can also give an intuitive explanation for the superfluid
shell state. Note that except for the deep BEC side with a very
strong coupling, it is always more favorable for the fermions to
pair up when the mismatch of the Fermi surfaces of the two
components becomes smaller. As one decreases the chemical
potential $\mu _{\mathbf{r}}$ by moving out from the trap center,
for non-interacting fermions the radius of the Fermi surface in
the momentum space $k_{F}^{\sigma }\left( \widetilde{r}\right) $
decreases as $k_{F}^{\sigma
}\left( \widetilde{r}\right) =\sqrt{2m_{\sigma }\left( \mu _{\sigma }-\mathbf{\widetilde{r}%
}^{2}\right)} $ (in the standard unit) for the component $\sigma $. So, $%
k_{F}^{\sigma }\left( \widetilde{r}\right) $ for the heavy fermions
decreases faster with increasing $\widetilde{r}$. This qualitative
statement should be true also for interacting fermions as
interaction will not change the rough trend. Then, we can imagine
two situations as depicted in Figs. 2(b) and 2(c). If at the trap
center, the Fermi surface of the heavy fermions ($^{40}$K) has a
smaller radius $k_{F}^{\sigma }\left( \widetilde{r}=0\right) $ (Fig.
2(b)), the mismatch of the two Fermi surfaces only grows as one increases $%
\widetilde{r}$. Therefore the pairing superfluid, if any, can only
form a core at the center. On the other hand, if the Fermi surface
of the heavy fermions has a larger
radius $k_{F}^{\sigma }\left( \widetilde{r}=0\right) $ (which is the case when $%
^{40}$K are in excess), the mismatch of the Fermi surfaces is
minimized at an intermediate region (see Fig. 2(c)), so the
superfluid only forms near that region and thus takes the shape of
a spherical shell. This explains an important qualitative feature
of the phase diagram in Fig. 1. As one further moves to the BEC
side, the Fermi surface mismatch (and thus the above mechanism)
becomes less important, and finally becomes independent as to
which component is in excess. The superfluid then always forms a
core at the trap center where there is a larger atomic density
(Fig. 1(d)).

\begin{figure}[tbp]
\includegraphics[width=8cm,height=4cm]{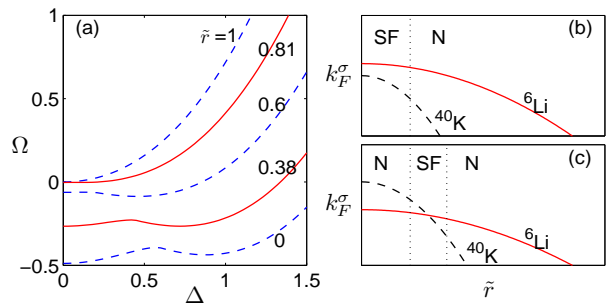}
\caption[Fig.2 ]{(color online) (a) The thermodynamical potential
$\Omega$ shown as a function of the order parameter $\Delta$ (both
in the unit of $E_F$) at different trap positions (different
chemical potentials) characterized by the normalized radius
$\widetilde{r}$. The other parameters are $p=0.2$, $T=0$, and
$(k_Fa_s)^{-1}=0$. The two solid curves bound a region
corresponding to the SF phase, where the global minimum of
$\Omega$ is at a nonzero $\Delta$.
(b) and (c) Schematic illustration of the radius of the Fermi surfaces $k_{F}^{%
\protect\sigma }$ for the two components as a function of the normalized
trap radius $\widetilde{r}$. In (b), one can only have a superfluid core at
the trap center, while in (c), one in general has a superfluid shell in the
intermediate region.}
\end{figure}

To detect the phase diagram of Fig. 1 in general and the
superfluid shell state in particular, one can measure the atomic
density profiles in the trap. The real-space density profiles for
the polarized fermi gas have been measured in several experiments
\cite{1,2,3,4}; in particular, the most recent one has shown how
to reconstruct the full density profile from the column integrated
signal \cite{4}. We calculate the density profiles for several
characteristic configurations of phase separation, and the results
are shown in Fig. 3. Figure 3(a) is for the resonance case with a
small population imbalance $p=0.2$ and at zero temperature. The
superfluid shell (where the densities for the two components are
equal) is clearly visible, which separates two normal regions.
From the inside normal state to the superfluid, the heavy (light)
fermion densities jump down (up), respectively. This jump is
consistent with the first-order phase transition picture
established from the thermodynamical potential $\Omega $ shown in
Fig. 2(a). From the superfluid shell to the outside normal
regions, the atomic densities drop continuously (consistent with
the second-order phase
transition picture in Fig. 2(a)). There is a small region of the NM$_{\text{%
Li}}$ state, and outside is a tail for the NP$_{\text{Li}}$
region. In Fig. 3(b), we show the finite temperature density
profiles ($T=0.1T_{F}$) with otherwise the same parameters as in
Fig. 3(a). The profiles get a bit more smooth (as one expects),
but the jump from the inside normal to the superfluid shell is
still clearly visible. Note that as pointed out in Ref.
\cite{wei1}, the densities are not equal any more for the BCS\
superfluid state at finite $T$ since quasiparticle excitations
carry population imbalance. Fig. 3(c) shows the density profiles
on the BCS\ side. There are jumps in the density profiles from the
superfluid shell to both the inside and the outside regions of
normal states. Fig. 3(d) shows the profiles on the BEC side, still
with a superfluid shell, but there is no normal region outside the
shell any more (no tails with unequal densities). On the BEC\
side, there is also a region of the BP1 state at appropriate
population imbalance. The BP1 state is hard to be directly seen
from the real-space density profile, but can be unambiguously
signaled with the detection of the momentum-space profile of the
minority component \cite{wei2}.

\begin{figure}[tbp]
\includegraphics[width=8cm,height=5cm]{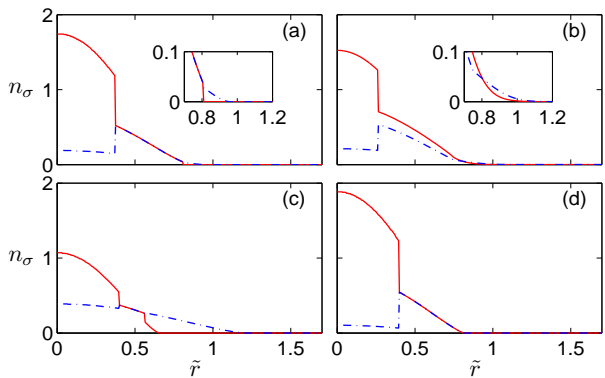}
\caption[Fig.3]{(color online) The atomic number densities
$n_{\protect\sigma}$ (in the unit of $n_F \equiv k_F^3
/(3\protect\pi^2)$) shown versus the normalized
trap radius $\widetilde{r}$. The solid (dashed) curves are for the $^{40}$K (%
$^{6}$Li) atoms, respectively. The inserts of (a) and (b) show the
amplified tails of the density profiles. The other parameters are:
(a)$T=0$, resonance,
and $p=0.2$, (b) $T=0.1 T_F$, resonance, and $p=0.2$, (c) $T=0$, $%
(k_Fa_s)^{-1}=-1$ (BCS side), $p=-0.4$, (d) $T=0$, $(k_Fa_s)^{-1}=0.2$ (BEC
side), $p=0.3$. }
\end{figure}

In summary, we have mapped out the phase diagram of a strongly
interacting fermion gas in a trap with both mass and population
imbalance for the two spin components. As a remarkable feature,
the superfluid forms a spherical shell in the case of mass
mismatch. We attribute this phenomenon to a combined effect of the
mass imbalance and the trapping potential. We also show that the
superfluid shell should be clearly visible in experiments through
explicit calculations of the atomic density profiles. The
superfluid shell is not simply connected in space, so it should
have nontrivial vortex properties under rotation of the trap,
which is an interesting subject for future theoretical and
experimental investigations.

This work was supported by the NSF awards (0431476), the ARDA under ARO
contracts, and the A. P. Sloan Fellowship.

\end{document}